\documentclass[proceedings]{JHEP} % 10pt is ignored

\conference{Trieste Meeting of the TMR Network on Physics beyond the
SM}

\title{Non-Supersymmetric Open String Vacua \footnote{Preprint CPHT-PC722.0799}
}

\author{Carlo Angelantonj\\
Centre de Physique Theorique, 
Ecole Polytechnique, \\
91128 Palaiseau CEDEX, France\\
E-mail: \email{angelant@cpht.polytechnique.fr}}

\abstract{We review the construction of non-supersymmetric open
string vacua in various dimensions. They can be obtained either
projecting the (compactified) non-supersymmetric 0B theory, or
applying the Scherk-Schwarz mechanism to open strings. Generically,
these vacua generate a non-vanishing cosmological constant. However,
one can construct particular kinds of Scherk-Schwarz compactifications
with vanishing cosmological constant, at least for low orders,
based on asymmetric orbifolds. A
generic feature of these models is that supersymmetry remains
unbroken on the branes at all mass level, while it is broken in the
bulk in a way that preserves Fermi-Bose degeneracy at each mass level
in the perturbative string spectrum}

\begin{document} 

\section{Introduction}

Motivated by the AdS/CFT correspondence conjecture \cite{malda}, 
in the last year
we have witnessed to an increasing interest in non-supersymmetric
string theories. On the one hand, the type 0B theory has been proposed
as the gravitational description (on some appropriate background) of
non-supersymmetric gauge theories on D3-branes \cite{kleb}. 
On the other hand, the
presence of non-supersym\-metric conformal fixed lines has suggested the
existence of non-supersymmetric string vacua with vanishing
cosmological constant \cite{ksold}. Since type 0 theories have only bosonic
excitations, this last scenario requires non-standard orbifold
compactifications of type II superstrings \cite{kks}. 

Consistent non-supersymmetric vacuum configurations in various
dimensions have been recently constructed in different contexts.
In the Field Theory
limit the Scherk-Schwarz mechanism \cite{ss} provides an elegant
realization of supersymmetry breaking by compactification. The higher
dimensional fields are single valued on the internal manifold up to a
(quantized) $R$-symmetry transformation. As a result, the scale
of supersymmetry breaking is quantized in units of the inverse size
of the internal
manifold. This mechanism has been extended to the full
perturbative spectrum in models of oriented closed strings \cite{sscs}
and in the corresponding open descendants \cite{ads}. 

It has long been known that, besides the five
supersymmetric strings, there are a number of non-supersymmetric
theories in ten dimensions \cite{nonsusy}. It was shown in \cite{bs}
that the type 0A and 0B theories allow the construction of open
descendants following \cite{cargese}. Typically, these theories
include in their spectrum tachyonic modes (both in the closed and
in the open sector) and thus have often
been regarded as toy models.
Recently, it has been shown that tachyons in the open sector can
be stabilized \cite{sen} and then do not represent an inconsistency
of the theory. Nevertheless, in order to study consistent string
vacua one should get rid of all tachyonic modes. 
To this end, one can use the crosscap constraint \cite{ccon}
to change the world-sheet projection, $\Omega$, in the closed
unoriented sector and remove all tachyonic modes both in the closed
and open unoriented sectors \cite{sag,zerob,bfl,bk} .

Typically, these two different approaches to the construction of
non-supersymmetric string vacua all generate a non-vanishing
cosmological constant at one loop. Thus, in order to describe the
four-dimensional low-energy world one should find mechanisms that
break supersymmetry without generating any cosmological constant
at least on suitable branes. 
These kinds of compactifications do exist and are related to asymmetric
orbifolds \cite{kks}. The main idea consists finding particular
kinds of projections that break supersymmetry only in the
(anti-)holomor\-phic sector, thus ensuring the vanishing of the
(one-loop) partition function. When 
projected with respect to the combined action of all the orbifold
generators, the resulting model is non-supersymmetric, but has an equal
number of fermionic and bosonic degrees of freedom at each mass level. 

This paper is organized as follows. In Section 2 we review some basic
facts about the construction of open descendants, and in particular
discuss the crosscap constraint. In Section 3 we discuss the
construction of tachyon-free open descendants of the 0B theory in six
and four dimensions. In Section 4 we present non-supersymmetric vacuum
configurations of open strings with vanishing cosmological constant.
These are open descendants of the model introduced in \cite{harv} .

\section{The crosscap constraint}

Let us review some known facts about the construction
of open descendants. The starting point is a theory of oriented closed
strings invariant under the world-sheet parity $\Omega$. For
example, one can consider type IIB or type 0 theories in $D=10$, or
any $d$-dimensional theory whose torus partition function
\begin{equation}
{\cal T} = \sum_{a,b} \chi_a N_{ab} \bar \chi _b 
\nonumber
\end{equation}
is defined through a GSO projection $N_{ab}$ compatible with
world-sheet parity. Then the $\Omega$-projection halves
the torus amplitude ${\cal T}$ and adds to it the contribution of
the Klein bottle. Actually, one has the freedom to project
onto states invariant under a modified world-sheet
symmetry $\Omega ' = \Omega {\cal J}$, where ${\cal J}$ is some
automorphism of the parent theory. Typically, ${\cal J}$ corresponds
to a different GSO projection in the parent theory that manifests
itself in a different modular invariant combination of characters (the
matrix $N_{ab}$) in the torus partition function. However, there are
cases in which ${\cal J}$ reflects an ambiguity in defining the Klein
bottle amplitude associated to a given GSO projection of the closed
oriented string.  In this case, they represent exhotic solutions to the
{\it crosscap constraint} \cite{ccon}.

The action of ${\cal J}$ on a given state $|a\rangle$ associated to
the character $\chi_a$ appearing diagonally in ${\cal T}$ 
translates into a sign $\sigma_a$ in the
corresponding contribution to the Klein bottle amplitude
\begin{equation}
{\cal K} = {\textstyle{1\over 2}} \sum_a \sigma_a K_a \chi_a
\nonumber
\end{equation}
consistently with supersymmetry (if present) and with the crosscap
constraint
\begin{equation}
\sigma_a \sigma_b \sigma_c = +1 \quad {\rm if } \quad N_{abc} \not= 0 \,,
\label{contsraint}
\end{equation}
where the tensor $N_{abc}$ encodes the fusion rules of the states
$|a\rangle$. 

An interesting example is the compactification of type IIB on a circle
of radius $R$. In this case, the $\Omega$-projection is accompained by
a $Z_2$ shift on the compact coordinate and then the resulting Klein
bottle amplitude
\begin{equation}
{\cal K} = {\textstyle{1\over 2}} (V_8 - S_8) \sum_m (-)^m q^{(m/R)^2}
\label{nostrings}
\end{equation}
introduces minus signs for the odd momentum states  
\cite{noopen,gepner}. Then, an $S$ modular transformation reveals that
there are no contributions from massless states to the transverse
channel that, as a result, does not develop any tadpole for massless
unphysical states. One is not allowed to introduce D-branes,
and the theory does not contain open strings.

More examples of generalized $\Omega '$-projections are offered by the
type 0B theory. One has four distinct choices, namely the standard
$\Omega$-projection and three additional $\Omega {\cal
J}$-projections, with ${\cal J} = \delta$ (the $Z_2$ shift of the 
previous example), ${\cal J} = (-)^{F_{{\rm R}}}$ and ${\cal J} =
(-)^{G_{{\rm R}}}$, where $F_{{\rm R}}$ ($G_{{\rm R}}$) is the
space-time (world-sheet) fermion number operator. In $D=10$ these
different projections have been analyzed in \cite{sag}. Particularly
interesting is the ${\cal J} = (-)^{G_{{\rm R}}}$ case, that 
results in a non-supersymmetric non-tachyonic model.
The non-perturbative consistency of these models has been recently
addressed in \cite{zerodual}. Their conclusion is that 
the $\Omega (-)^{F_{{\rm R}}}$ projection is ruled out
since in the decompactification limit it does not lift to any symmetry
of M-theory. This leaves only the standard $\Omega$ projection and the
non-tachyonic one.

\section{Open descendants of the 0B string}

The starting point is the 0B string in $D=10$. 
It can be constructed as an orbifold of type IIB, where the symmetry
to be gauged is $(-)^{F_{{\rm L}}+F_{{\rm R}}}$.
Obviously, this projection removes all the (space-time)
fermions and introduces, in the twisted sector, a tachyon and
additional RR states. In terms of characters of affine ${\rm SO} (8)$
at level one, the massless and massive excitations of the 0B string
can then be collected in the torus partition function
\begin{equation}
{\cal T} = |O_8|^2 + |V_8|^2 + |S_8|^2 + |C_8|^2 \,,
\nonumber
\end{equation}
aside from transverse bosonic modes.
Due to the presence of the tachyon, the theory is unstable unless one
goes off-criticality, and the {\it vev} of the tachyon is chosen to precisely
compensate the deficit of central charge \cite{gab}.
Ten-dimensional stable non-tachyonic vacuum configurations can indeed
be constructed modding out type 0B by the combination $\Omega
(-)^{G_{{\rm R}}}$ \cite{sag}. Since the Klein bottle amplitude now
antisymmetrizes all the sta\-tes associated to the $O_8$ character, this
projection removes the ta\-chyon from the closed spectrum. Typically, this is
not sufficient to guarantee the stability of the open descendants. A
new complex tachyon in the bi-fundamental of a gauge group ${\rm U}
(N) \otimes {\rm U} (32+N)$ appears in the open unoriented sector,
where each unitary group comes from the two different D9-branes that exist 
in type 0B. However, tadpole conditions do not fix the number of D9
and, say, D$9'$-branes separately. As a result, one is free to choose
$N=0$, finally obtaining  a non-tachyonic model with fixed gauge
group. It should be noted that the consistency of the construction
forces one to relax the dilaton tadpole. For 0B theories this does not
cause any problem, since NS and R fields are no longer
related by space-time supersymmetry. 
The tadpole of the dilaton, or of any other
physical field, can then be disposed by the Fishler-Susskind mechanism
\cite{fs}. 

One can now try to generalize this construction  to $d$-dimensional
vacua. Toroidal compactifications can be treated precisely in the same
way, since they do not introduce new structures.  On the contrary,
interesting new features are present in orbifold compactifications. Since
the starting theory is not supersymmetric, one can study both
``supersymmetric'' orbifolds \cite{zerob,bfl} and ``non-supersymmetric''
ones \cite{bk} \footnote{By ``supersymmetric'' orbifolds here we mean
orbifold compactifications that would give supersymmetric vacua if one
started from type II, type I or (supersymmetric) heterotic strings.}.
For ``supersymmetric'' $T^4 /Z_N$ and $T^6 /Z_N$ an exhaustive
analysis shows that, for generic $N$, the untwisted sector includes a
tachyon, directly related to the ten-dimensional tachyon of 0B,
invariant under the action of $Z_N$. Typically, one finds 
additional (complex)
tachyons in the twisted sectors. However, these can not be
projected out in the open descendant by any $\Omega {\cal J}$
projection at generic (non-rational) points in the moduli space 
since, for a geometric action of the orbifold generators,
the associated partition function belongs to the family of charged
conjugate modular invariants. This means that states in the
$\theta$-twisted sector are pai\-red with states in the
$\theta^{-1}$-twisted one, and the net number of tachyons is then simply
halved by any $\Omega '$-projection. An exception are the orbifolds
$T^6 /Z_3$ and $T^4/Z_2$, since they do not generate tachyonic modes
in the twisted sectors.

The $T^6 /Z_3$ orbifold \cite{zerob,bfl} resembles the one studied in
\cite{chiral} aside from the fact that now one is starting from the 0B
 theory. The low-lying excitations of the resulting theory comprise a
tachyon, the metric tensor, two abelian vectors and sixty scalars from
the untwisted sector and $162=27\times 6$ scalars from the 27 fixed
points in the twisted sectors. The only states that are effectively
left-right (anti-)symmetric are the tachyonic vacuum, $g_{\mu\nu}$,
the two $A_\mu$ and two scalars related to the dimensional reduction
of the dilaton and of the (dualized) NS-NS antisymmetric tensor. Then,
the standard Klein bottle amplitude would project out the two vectors
and the $B_{\mu\nu}$, while keeping the tachyon. On the contrary, 
combining the
world-sheet parity with the world-sheet fermion number, $\Omega
(-)^{G_{{\rm R}}}$, has the virtue of removing the tachyon from the
spectrum while keeping one of the two vectors, consistently with the
crosscap constraint. The remaining massless states are simply halved
by any $\Omega '$-projection, since their holomorphic and
antiholomorphic parts come from different sectors. Summarizing, the
non tachyonic spectrum comprises the graviton, an abelian vector and
111 scalars.

Moving to the open unoriented sector, also in this case we see a
doubling of the gauge group. This reflects the doubling of the RR
sector in the 0B theory, that introduces two different kinds of
D9-branes. In addition, the $(-)^{G_{{\rm R}}}$ projection enhances
the ${\rm SO} (n)$ factors that one would naively expect for a $Z_3$
orbifold \cite{chiral} to ${\rm U}(n)$ groups. Thus, tadpole 
cancellations\footnote{Also in this case, as in the $T^4/Z_2$
case that we will discuss later on, the Fishler-Susskind mechanism
is to be invoked to cure the uncancelled dilaton tadpole.}
partially fix the form of the Chan-Paton gauge group 
\begin{eqnarray}
G_{{\rm CP}} &=& \left[ {\rm U} (n) \otimes {\rm U} (m)^2
\right]_{99} \otimes 
\nonumber\\
& & \otimes \left[{\rm U} (8+n) \otimes {\rm U} (12+m) ^2
 \right]_{9'9'} \,.
\nonumber
\end{eqnarray}
while leaving the total dimension arbitrary.
As for the ten-dimensional case, one finds in the spectrum a complex
tachyon in the $99'$ sector.
It is then evident that the choice $n=m=0$ leads to an open sector
free of tachyons with a ${\rm U} (8) \otimes {\rm U} (12)^2$ gauge
group. The remaining (charged) massless excitations comprise three
scalars in the $[ (\overline{{\bf 8}} , {\bf 12}, {\bf
1}) \oplus ({\bf 8} , {\bf 1} , \overline{{\bf 12}}) \oplus ({\bf 1} ,
\overline{{\bf 12}} , {\bf 12} ) \oplus {\rm c.c} ]$ representations, one Dirac
fermion in the $({\bf 28} ,$ ${\bf 1}, {\bf 1})\oplus ({\bf 1} , {\bf 12}
, {\bf 12})$ representations, and, finally, three chiral fermions in
the $(\overline{{\bf 8}} , {\bf 1} , \overline{{\bf 12}}) \oplus ({\bf
8} , {\bf 12} , {\bf 1} ) \oplus ({\bf 1} ,\overline{{\bf 66}} ,$ ${\bf 1})
\oplus ({\bf 1} , {\bf 1} ,{\bf 66})$ representations
\cite{zerob,bfl}. The vanishing of the irreducible part of the 
gauge anomaly is a consequence of the cancellation of twisted 
tadpoles \cite{alda}.

Let us now turn to the $T^4 /Z_2$ orbifold. As one would expect, here we
have a richer structure, since the orbifold group includes an
element that squares to the identity (the generator itself). As a
result D5-branes are also present \cite{ps,gp}. Moreover, the projection
$\Omega (-)^{G_{{\rm R}}}$ introduces minus signs in the twisted sector
of the Klein bottle amplitude, consistently with the fact that
the fusion of an untwisted state with a twisted one gives another state that
belongs to the twisted sector: the crosscap constraint then
forces one to introduce further signs in the twisted sector.  
As we have already discussed, the inclusion of the world-sheet fermion
number in the $\Omega '$-projection removes the tachyon from the
closed unoriented sector. The massless states that survive the
projection give a chiral spectrum consisting of the metric tensor, 4
antiself-dual 2-forms, 20 self-dual 2-forms and 99 scalars, while the
open unoriented sector contains two pairs of D9 and D5-branes with a
Chan-Paton gauge group 
\begin{eqnarray}
G_{{\rm CP}} &=& \left[ {\rm U} (n) _{99} \otimes {\rm U} (16+n)_{9'9'}
\right] \otimes 
\nonumber \\
& & \otimes \left[ {\rm U} (m) _{55} \otimes {\rm U} (16+m)_{5'5'}
\right]
\end{eqnarray}
The open tachyons correspond to open strings stretched between  
D9 and D$9'$ branes and D5 and D$5'$ branes. It is then evident that
if one sets $n$ and $m$ to zero, consistently with tadpole conditions,
the resulting model is free of any closed and open tachyons. The remaining
massless scalars and spinors charged under the Chan-Paton gauge group
contribute to the cancellation of the irreducible part of the
anomaly polynomial, while a generalized Green-Schwarz mechanism is at
work to cancel the residual reducible anomaly \cite{ggs}. 

Actually, this model allows two different spectra of
charged states \cite{zerob,bfl}. This is due to the possibility of
adding discrete Wilson lines in the M\"obius amplitude, {\it i.e.}
relative phases between holes and crosscaps \cite{bs}. This results into a
different $P$ transformation, and, thus, in a different M\"obius
vacuum amplitude. Precisely the inclusion of discrete Wilson lines
in $D=6$
allowed the authors of \cite{bs} to construct the ${\rm U} (16) \otimes
{\rm U} (16)$ model, rediscovered after some years in \cite{gp}.
Different $T^4/Z_2$ type 0B descendants neatly 
can be constructed at rational points in
the internal lattice. For example, starting from the SO(8) lattice as
in \cite{bs,gepner} new massless scalars collapse to zero mass,
in a way reminiscent of
what happens in the bosonic and heterotic strings,
for which one has symmetry enhancement at special points in the moduli
space. Moreover, one has several choices of inequivalent sign
assignements for the Klein bottle amplitude, that lead to different
non-tachyonic vacuum configurations. Due to the presence of an
antisymmetric tensor background in the SO(8) lattice, one expects
Chan-Paton gauge groups of reduced rank \cite{toroidal} 
and additional self-dual 
2-forms related to a different Klein bottle projection in the twisted
sector \cite{new}.

Compactifications on the simplest instance of ``non-supersymmetric''
orbifolds have been recently considered by Blumenhagen and
Kumar \cite{bk}. They studied the $T^6 /Z_2$ case, where all internal 
coordinates are reversed under the $Z_2$. Typically, this particular
projection does not satisfy the constraints of modular invariance for
the superstring. It acts as a $Z_4$ action on the Ramond sector, and
introduces phases in the NS-R and R-NS sectors. However,
the action on the type 0B is perfectly
consistent. One can project further the theory by 
$\Omega (-)^{G_{{\rm R}}}$ to get non-tachyonic four-dimensional 
open string vacua. The result is similar to the 6D one
previously discussed, but with D3-branes instead of D5-branes.

\section{Non-supersymmetric vacua with vanishing cosmological
constant}

Non-supersymmetric vacua with a vanishing cosmological constant can
be obtained as asymmetric orbifolds of type II and/or type I
superstrings. The simplest instance of these models can be obtained
considering the generators \cite{harv}
\begin{eqnarray}
\!\!\!\!\!\!\!\! 
f &=& [(-1^4,1;1^5),(0^4,v_{{\rm L}} ; \delta^4 , v_{{\rm R}} ) ,
(-)^{F_{{\rm R}}} ] \,,
\nonumber \\
\!\!\!\!\!\!\!\!
g &=& [(1^5 ; -1^4 , 1) , (\delta^4 , w_{{\rm L}} ; 0^4 , w_{{\rm R}}
) , (-)^{F_{{\rm L}}} ] \,, \label{generators}
\end{eqnarray}
that define a non-abelian space group orbifold $S$ \cite{dhvw}.
Here the first entry inside the square brackets denotes rotations,
the second denotes shifts on the internal compactification lattice while the
third entry denotes projections on states with even left (right) 
space-time fermion number. A
semicolon separates holomorphic and antiholomorphic coordinates.
The asymmetric nature of the orbifold requires that the internal
four-dimensional lattice splits into a product of four circles
with self-dual radius $R=\sqrt{\alpha '}$. Level matching requires
$\delta$ be a shift by $R/2$, while no further constraints are imposed
on the radius of the fifth coordinate, for which the shifts $v_{{\rm L,R}}
= w_{{\rm R,L}}$ act as $A_2$ shifts in the notation of \cite{vw}.

Due to the presence of $(-)^{F_{{\rm L,R}}}$, the $g$ $(f)$ generator
projects out all gravitini coming from the (anti)holomorphic sector,
and therefore the combined action of $f$ and $g$ breaks supersymmetry
completely, while ensuring the vanishing of the one-loop
contribution to the cosmological constant. In \cite{ks} it was argued
that higher order perturbative corrections to the cosmological
constant vanish as well. There are, however, non-perturbative
contributions originating from wrapped D-branes that can be studied
perturbatively on the dual heterotic model \cite{harv} or computing
the D-brane spectra using boundary-state techniques \cite{kors}. 

In order to construct the orbifold (\ref{generators}), it
is simpler to restrict onself to the abelian point group $\overline{P}$
\cite{dhvw}, thus defining a new compactification lattice
$\Gamma_{4,4} ({\rm SO} (8)) \oplus \Gamma_{1,1} (R)$. In this case,
$\overline{P}$ reduces to an abelian $Z_2 \otimes Z_2$ orbifold. 
The resulting one-loop partition function involves only
the modular orbit generated by the untwisted sector. This fact has two
different interpretations in the space group and in the point group
approaches to the orbifold. In the former, it is due to the fact
that the path integral receives contributions only from commuting
spin structures \cite{dhvw}, whereas in the latter the
disconnected orbit vanishes due to the simultaneous action of shifts
and rotations \cite{kk}.

Invariance under world-sheet parity forces one to start with the type IIB
superstring modded out by the elements (\ref{generators}).
Using the characters of affine ${\rm
SO} (2n)$ at level one, the massless contributions to the torus
amplitude read
\begin{eqnarray}
{\cal T}_{{\rm untw}} &\sim& |V_4 O_4|^2 + |S_4 S_4 |^2 +
\nonumber \\
& & - (O_4 V_4 )(
\bar C _4 \bar C _4 ) - (C_4 C_4 )(\bar O_4 \bar V_4 ) \,,
\nonumber \\
{\cal T}_{fg{\rm -tw}} &\sim& 8\, |O_4 S_4 - C_4 O_4 |^2 \,,
\nonumber
\end{eqnarray}
that translate into the following five-dimensional field content: the
metric tensor, seven abelian vectors, six scalars and eight fermions
from the untwisted sector, and eight vectors, forty scalars and
sixteen fermions from the $fg$-twisted sector. Due to the presence of
the shifts, the $f$ and $g$-twisted sectors are massive, so
that no massless gravitini originate from them.

The construction of the open descendants requires
suitable contributions from the Klein bottle, annulus and
M\"obius strip world-sheet topologies \cite{cargese, bs}. In the
following we will just describe the main features of the model and we
refer the interested reader to \cite{aaf} for
details\footnote{See also \cite{bg} for a different approach to the
construction of the D-brane structure for the asymmetric orbifold
(\ref{generators}).}. 
At the massless level one finds the following contributions:
\begin{equation}
{\cal K} \sim (V_4 O_4 - S_4 S_4 ) + (6-2)\, (O_4 S_4 - C_4 O_4 ) 
\label{klein}
\end{equation}
from the Klein bottle amplitude,
\begin{eqnarray}
{\cal A} &\sim& 2 M \bar M \, (V_4 O_4 - S_4 S_4 ) + 
\nonumber \\
& & + (M^2 + \bar M ^2 )
(O_4 V_4 - C_4 C_4 )
\label{annulus}
\end{eqnarray}
from the annulus amplitude, and, finally,
\begin{equation}
{\cal M} \sim - (M+\bar M ) (\hat O _4 \hat V_4 - \hat C_4 \hat C_4 ) 
\label{moebius}
\end{equation}
from the M\"obius amplitude. Since the action of the generators $f$ and $g$
is equivalent to a T-duality transformation, the
open unoriented sector involves only one kind of charges, 
a linear combination of (wrapped) D9 and D5-branes. In the Klein
bottle amplitude we have explicitly written how the
$\Omega$-projection acts on the $fg$-twisted sector. Not all the
contributions are treated in the same way. $\Omega$
(anti-)symmetri\-zes  
the (R-R) NS-NS sector of six ($f$-invariant
combinations of) fixed points , and (anti-)symmetrizes
the (NS-NS) R-R sector of the remaining two
fixed points. This is
due to the presence of a non-vanishing $B_{ab}$ background in the
${\rm SO} (8)$ lattice \cite{orbtor,new}.

The cancellation of tadpoles of unphysical massless states in the
transverse channel then fixes the size of the Chan-Paton gauge
group, so that $M=\bar M = 8$. 
The rank reduction is due both to the presence of a
non-vanishing background for the NS-NS antisymmetric tensor in the
${\rm SO} (8)$ lattice \cite{toroidal,orbtor,new} and to the
identification of (wrap\-ped) D9 and D5-branes under the action of the
orbifold generators (\ref{generators}). The
massless closed unoriented spectrum
comprises a graviton, 4 abelian vectors, 31 scalars and 12
fermions. The massless unoriented open
sector is by itself supersymmetric, and comprises an ${\cal N} =2$
vector multiplet with gauge group ${\rm U} (8)$, as well as a
hypermultiplet in the representations ${\bf 28} \oplus
\overline{{\bf 28}}$. Moreover, the amplitudes ${\cal K}$, ${\cal
A}$, and ${\cal M}$ are supersymmetric at each mass level, 
compatibly with the vanishing of the (one-loop) cosmological constant. This
has to be contrasted with what usually happens in `M-theory
breaking' of type I models \cite{ads} and in the heterotic dual of
\cite{harv,aaf}, where supersymmetry is realized in the ``gauge
sector'' only at the massless level. 

In conclusion, one can study the limiting behavior of the type I
model for large and small radius.  For $R\to 0$ new uncancelled tadpoles
arise in the transverse channels, 
thus inducing linear divergences in the five-dimensional
gauge theory on the branes \cite{ab}. On the contrary, as 
$R\to \infty$ the action of the
$f$ and $g$ generators is trivialized, while the sector
projected and/or twisted by the combined $fg$ action survives. The
decompactification limit then corresponds to a $T^4 /Z_2$ orbifold.
Due to the presence of a non-vanishing flux for the NS-NS $B_{ab}$,
the type I compactification comprises at massless level additional
tensor multiplets and has a Chan-Paton group with reduced rank. One finds
a $D=6$ ${\cal N}=(1,0)$ supergravity multiplet coupled to 5
tensor multiplets and 16 hypermultiplets from the closed unoriented
sector together with a vector multiplet in the adjoint representation
of ${\rm U} (8) \otimes {\rm U} (8)$ and charged hypermultiplets in
the representations $({\bf 28}\oplus \overline{{\bf 28}} , {\bf 1})
\oplus ({\bf 1},{\bf 28}\oplus \overline{{\bf 28}}) \oplus ({\bf 8},
\overline{{\bf 8}})\oplus (\overline{{\bf 8}} , {\bf 8})$ from the
open unoriented sector, one of the models of \cite{bs,gepner}.
The presence of additional
tensor multiplets calls for a generalized Green-Schwarz mechanism for
the cancellation of the residual anomaly \cite{ggs}.

\acknowledgments 
It is a pleasure to thank I. Antoniadis and K. F\"orger for a
pleasant and stimulating collaboration. I would also like to
thank A. Sagnotti for stimulating discussions. 
Research supported in part by the EEC  TMR contract
ERBFMRX-CT96-0090.


\begin{thebibliography}{99}

\bibitem{malda}{J. Maldacena, \atmp{2}{1998}{231}.}

\bibitem{kleb}{I.R. Klebanov and A.A. Tseytlin,
\npb{546}{1999}{155}; \jhep{03}{1999}{015}; \npb{547}{1999}{143};\\
J.A. Minahan, \jhep{04}{1999}{007};\\ A. Armoni and B. Kol,
`Non-supersymmetric large $N$ gauge theories from type 0 brane
configurations', \hepth{9906081};\\
S. Seki, `Baryon configurations in the UV and IR regions of type 0
string theory', \hepth{9906210};\\
I.R. Klebanov, `Tachyon stabilization in the Ads/CFT correspondence',
\hepth{9906220}.}

\bibitem{ksold}{S. Kachru and E. Silverstein, \prl{80}{1998}{4855}.}

\bibitem{kks}{S. Kachru, J. Kumar and E. Silverstein,
\prd{59}{1999}{106004}.}

\bibitem{ss}{J. Scherk and J.H. Schwarz, \npb{153}{1979}{61}.}

\bibitem{sscs}{R. Rohm, \npb{237}{1984}{553};\\
C. Kounnas and M. Porrati, \npb{310}{1988}{355}; \\
S. Ferrara, C. Kounnas, M. Porrati and F. Zwirner,
\npb{318}{1989}{75};\\
C. Kounnas and B. Rostand, \npb{341}{1990}{641};\\
I. Antoniadis and C. Kounnas, \plb{261}{1991}{369};\\
E. Kiritsis and C. Kounnas, \npb{503}{1997}{117}.}

\bibitem{ads}{I. Antoniadis, E. Dudas and A. Sagnotti,
\npb{544}{1999}{469};\\
I. Antoniadis, G. D'Appollonio, E. Dudas and A. Sagnotti, `Partial
breaking of supersymmetry, open strings and M-theory', \hepth{9812118}.}

\bibitem{nonsusy}{L.J. Dixon and J.A. Harvey, \npb{274}{1986}{93};\\
N. Seiberg and E. Witten, \npb{276}{1986}{272};\\ L.
Alvarez-Gaum\'e, P. Ginsparg, G. Moore and C. Vafa,
\plb{171}{1986}{155}.}

\bibitem{bs}{M. Bianchi and A. Sagnotti, \plb{247}{1990}{517};
\npb{361}{1991}{519}.}

\bibitem{cargese}{A. Sagnotti, in Carg\`ese `87, Non-perturbative
Quantum field Theory, G. Mack, et al. (Eds.), Pergamon Press, Oxford,
1988, p. 521.}

\bibitem{sen}{A. Sen, \jhep{06}{1998}{007}; \jhep{08}{1998}{010};
\jhep{08}{1998}{012}; \jhep{09}{1998}{023}; \jhep{10}{1998}{021};
\jhep{12}{1998}{021};\\
O. Bergman and M.R. Gaberdiel, \plb{441}{1998}{133};
\jhep{03}{1999}{013};\\
E. Witten, \jhep{12}{1998}{019};\\
M. Frau, L. Gallot, A. Lerda, P. Strigazzi, `Stable non-BPS D-branes
in type I string theory', \hepth{9903123}.}

\bibitem{ccon}{D. Fioravanti, G. Pradisi and A. Sagnotti,
\plb{321}{1994}{349};\\
G. Pradisi, A. Sagnotti and Ya.S. Stanev, \plb{354}{1995}{279};
\plb{356}{1995}{230}; \plb{381}{1996}{97}.}

\bibitem{sag}{A. Sagnotti, `Some surprises of open-string theories',
\hepth{9509080}; `Surprises in open-string perturbation theory',
\hepth{9702093}.}

\bibitem{zerob}{C. Angelantonj, \plb{444}{1998}{309}.}

\bibitem{bfl}{R. Blumenhagen, A. Font and D. L\"ust, `Tachyon-free
orientifolds of type 0B strings in various dimensions',
\hepth{9904069}.}

\bibitem{bk}{R. Blumehagen and A. Kumar, 
`A note on orientifolds and dualities of
type 0B string theory', \hepth{9906234}.}

\bibitem{fs}{W. Fischler and L. Susskind, \plb{171}{1986}{383};
\plb{173}{1986}{262}.}

\bibitem{harv}{J.A. Harvey, \prd{59}{1999}{26002}.}

\bibitem{noopen}{A. Dabholkar and J. Park, \npb{477}{1996}{701}.}

\bibitem{gepner}{C. Angelantonj, M. Bianchi, G. Pradisi, A. Sagnotti
and Ya.S. Stanev, \plb{387}{1996}{743}.}

\bibitem{zerodual}{O. Bergman and M.R. Gaberdiel, `Dualities of type 0
strings', \hepth{9906055}.}

\bibitem{gab}{G. Ferretti and D. Martelli, `On the construction of
gauge theories from noncritical type 0 strings', \hepth{9811208};\\
G. Ferretti, J. Kalkkinen and D. Martelli,
`Noncritical type 0 string theories and their field theory duals',
\hepth{9904013}.}

\bibitem{chiral}{C. Angelantonj, M. Bianchi, G. Pradisi, A. Sagnotti
and Ya.S. Stanev, \plb{385}{1996}{96}.}

\bibitem{alda}{J. Polchinski and Y. Cai, \npb{296}{1988}{91};\\
G. Aldazabal, D. Badagnani, L.E. Ib\'a\~{n}ez and
A.M. Uranga, `Tadpole versus anomaly cancellation in $D=4$, $D=6$
compact IIB orientifolds', \hepth{9904071}.}

\bibitem{ps}{G. Pradisi and A. Sagnotti, \plb{216}{1989}{59}.}

\bibitem{gp}{E. Gimon and J. Polchinski, \prd{54}{1996}{1667}.}

\bibitem{ggs}{A. Sagnotti, \plb{294}{1992}{196};\\
S. Ferrara, R. Minasian and A. Sagnotti, \npb{474}{1996}{323};\\
S. Ferrara, F. Riccioni and A. Sagnotti, \npb{519}{1998}{115};\\
F. Riccioni and A. Sagnotti, \plb{436}{1998}{298}.}

\bibitem{toroidal}{M. Bianchi, G. Pradisi and A. Sagnotti,
\npb{376}{1992}{365};\\
M. Bianchi, \npb{528}{1998}{73};\\
E. Witten, \jhep{02}{1998}{006}.}

\bibitem{new}{C. Angelantonj, in preparation.}

\bibitem{dhvw}{L. Dixon, J.A. Harvery, C. Vafa and E. Witten,
\npb{261}{1985}{678}; \npb{274}{1986}{285}.}

\bibitem{vw}{C. Vafa and E. Witten, \npps{46}{1996}{225}.}

\bibitem{ks}{S. Kachru and E. Silverstein, \jhep{11}{1998}{001}; 
\jhep{01}{1999}{004}.}

\bibitem{kors}{B. K\"ors, `D-brane spectra of nonsupersymmetric,
asymmetric orbifolds and nonperturbative contributions to the
cosmological constant', \hepth{9907007}.}

\bibitem{kk}{E. Kiritsis and C. Kounnas, \npb{503}{1997}{117}.}

\bibitem{aaf}{C. Angelantonj, I. Antoniadis, K. F\"orger,
`Non-supersymmetric type I strings with zero vacuum energy',
\hepth{9904092}.}

\bibitem{bg}{R. Blumenhagen and L. G\"orlich, \npb{551}{1999}{601}.} 

\bibitem{orbtor}{Z. Kakushadze, G. Shiu and S.-H.H. Tye,
\prd{58}{1998}{086001}.}

\bibitem{ab}{I. Antoniadis and C. Bachas, \plb{450}{1999}{83}.}











\end{thebibliography}
\end{document}